# Quantum Spectra of Hydrogen Atoms in Various Magnetic Fields with the Closed Orbit Theory *

PENG Liang-You(彭良友)[1], ZHANG Xian-Zhou(张现周)[1], RAO Jian-Guo(饶建国)[1,2]
[1]Wuhan Institute of Physics and Mathematics, Chinese Academy of Sciences, Wuhan 430071
[2]Department of Applied Mathematics and Theoretical Physics, The Queen's University of Belfast, Belfast BT7 1NN, UK

(Received 1 August 2001)

The quantum spectra of hydrogen atoms in various magnetic fields have been calculated with the closed orbit theory. The magnitude of the magnetic field decreases from 5.96 T to 0.56 T with a step of 0.6 T. We demonstrate schematically that the closed orbits disappear with the decrease of the magnitude of the magnetic field when the corresponding finite resolution of experiment is fixed. This may give us a good way to control the shape and the number of the closed orbits in the system, and thus to control where a peak should exist in the Fourier transformation of the quantum spectra.

PACS: 32. 70. Cs, 32. 60. +i

Rydberg atoms in external magnetic fields, as a real physical example of a nonintegrable system, have been proven to be ideally suitable for the study of quantum manifestations of classical chaos experimentally and theoretically.[1] Even the apparently simplest system, the hydrogen atom situated in a magnetic field, is neither separable nor integrable in the regime where the magnetic force is comparable with the Coulomb force when the hydrogen atom is excited close to the ionization threshold.[2] Since the discovery of quasi-Landau-type resonance in the spectra of barium[3] and hydrogen atoms,[4] various theories have been developed to interpret this exciting phenomenon. A decisive breakthrough for a semiclassical interpretation of structures in the photoabsorption cross section has been devised by Du and Delos;[5] this is known as *the closed orbit theory*. It has also been successfully used to explain the spectra of atoms in an external electric field.[6] Even for strontium atoms, which have large ionic cores, in an electric field the positions of most experimental recurrence peaks coincide well with the theory, with the strengths of the peaks modulated by a strong core scattering.[7] Also, closed orbits have been shown recently to play an important role in the interpretation of the photoexcitation scaled spectrum of the hydrogen atom in crossed magnetic and electric fields.[8]

However, theoretical calculations until now have exclusively dealt with a uniform magnetic field. Few calculations have been carried out with atoms in magnetic fields of various magnitudes. In this letter, we calculate the quantum spectra of hydrogen atoms in magnetic fields of ten different magnitudes using the closed orbit theory. The Hamiltonian of the system in a precession coordinate about the z-axis at the Larmor frequency (in the spherical coordinate) is

$$H = \frac{1}{2}\left(p_r^2 + \frac{p_\theta^2}{r^2} + \frac{L_z^2}{r^2\sin^2\theta}\right) - \frac{1}{r} + \frac{\gamma^2(t)}{8}r^2\sin^2\theta, \quad (1)$$

where $\gamma(t) = B(t)/B_0$, $B_0 = 2.35 \times 10^5$ T, and $L_z = m\hbar$ is the z component of angular momentum. The magnetic field changes with time according to

$$B(t) = 5.96 - 0.6t, \quad (2)$$

where $t$ takes the integers ranging from 0 to 9.

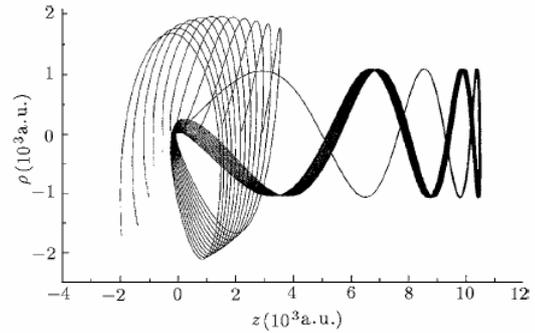

**Fig. 1.** A number of classical trajectories whose initial angles are centred around 25.70°, showing very intuitively the sensitive dependence to the initial conditions in classical chaos.

We will consider the transition from $2p_{m=0}$ to $m = 0, E \approx 0$. According to the closed orbit theory,[5] the observed oscillator strength can be written as a combination of a smooth background and a sum of sinusoidal oscillations

$$\overline{Df}(E) = Df_0(E) + \sum_n A_n(E) \cdot \sin\left(\int_0^E T_n(E')dE' + \alpha_n\right). \quad (3)$$

The first term is precisely the oscillator strength that would be reached without any external field. Each term in the summation of the second part corresponds to a closed orbit of the electron in the combined Coulomb and magnetic fields. $T_n(E)$ is the time taken for the electron to start from and end at the vicinity of the nucleus for the $n$th closed orbit. $A_n(E)$ is the amplitude of each oscillation and $\alpha_n$ is the phase constant for the same oscillation.

*Supported by the National Natural Science Foundation of China under Grant No. 19874071.
©2002 Chinese Physical Society and IOP Publishing Ltd

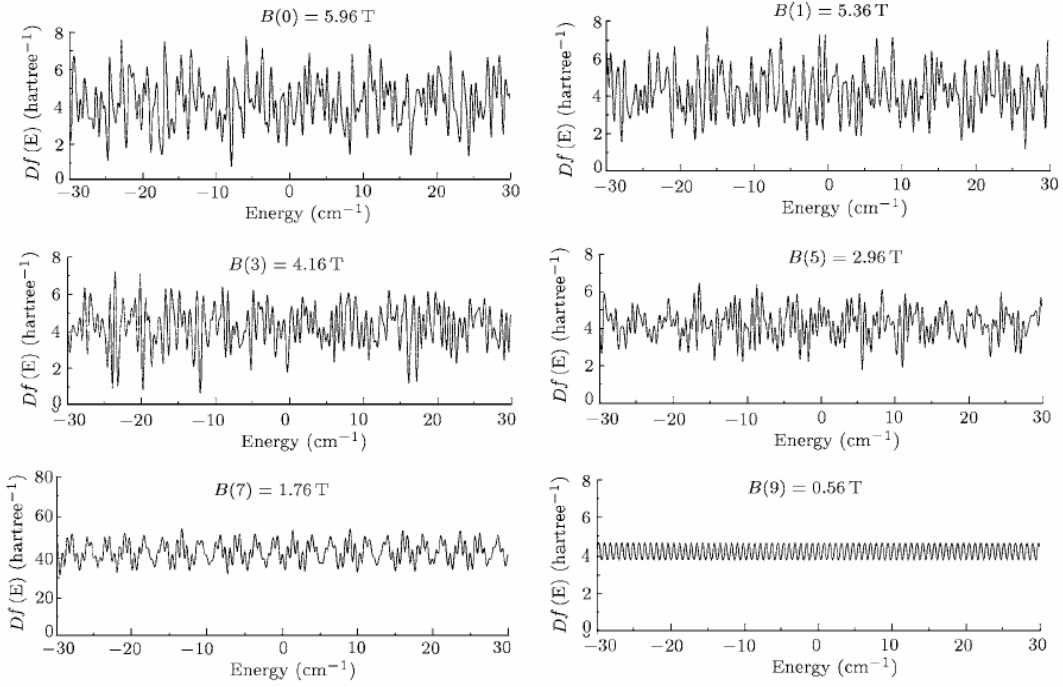

**Fig. 2.** Absorption spectra of hydrogen atoms in various magnetic fields. We see that the spectra become more and more simple when $B(t)$ decreases. The corresponding value of the magnetic field is indicated respectively. The photo energy (horizontal axis) increases from left to right.

The closed orbit theory is based on the assumption that the space is divided into two regions. In the region close to the nucleus, although the effect of the magnetic field can be neglected, a pure quantum description is required. However, in the region far from the nucleus, a classical description is sufficient. The electron will propagate outwards, following classical trajectories. Some of the trajectories will be turned back by the magnetic field to the vicinity of the nucleus, forming closed orbits. The two different descriptions are matched at a certain suitable boundary, $r_b$, which is taken to be $50a_0$ ($a_0$ is the Bohr radius) in our work.

It is easy to calculate the smooth part[5]

$$Df_0(E=0) = 4.2138\,\text{Hartree}^{-1}. \quad (4)$$

The oscillatory part is connected to some classical quantities of the closed orbits found in the system. $A_n(E)$ and $\alpha_n$ in Eq. (3) are calculated by evaluating

$$\begin{aligned}&A_n(E)\exp[i\alpha_n(E)]\\&= \frac{m_e}{\hbar^2}(E-E_i)2^{19/4}\pi^{3/2}r_b^{-1/4}[\sin\theta_i^n\sin\theta_f^n]^{1/2}A_2^n\\&\quad\cdot e^{-i\frac{3\pi}{4}}e^{i2(8r_b)^{1/2}}\exp[i(\frac{S_2^n}{\hbar}-\frac{1}{2}\pi\mu^n)]\wp(\theta_i^n)\wp(\theta_f^n),\end{aligned}\quad(5)$$

where

$$\wp(\theta) = \sum_{l_1}(-1)^{l_1}I(n,l,l_1)b^i_{l_1 m}Y_{l_1 m}(\theta,0). \quad (6)$$

In Eq. (5), $E$ and $E_i$ are the energies of the final and initial states, respectively; $\theta_i^n$ and $\theta_f^n$ are the initial and final angles of the $n$th closed orbit, respectively; $A_2^n$ and $S_2^n$ are the two-dimensional amplitude and classical action; and $\mu^n$ is the Maslov index of the closed orbit. The definitions of the quantities on the right-hand side of Eq. (6) are the same as those in Ref. [5].

We use the fifth-order Cash–Karp Runge–Kutta method with an estimation of the local truncation error using the embedded fourth-order method to integrate the motion equations. In order to search the accurately closed orbits and calculate the relevant classical quantities, we follow three steps in the computer codes, as follows. (1) We search those nearly closed orbits in the range of $(0°, 90°)$ instead of $(0°, 180°)$ because of the symmetry of the problem. In order to save computer time, we set the step size of the initial launching angle $\theta_i$ to be $0.1°$. It is recorded as a nearly closed orbit when the trajectory returns to the vicinity of the nucleus when $|r - r_b| < 1.0 a_0$ (certainly, we require $p_r < 0$). (2) We change the step size of $\theta_i$ to $0.01°$, then we search two nearly closed orbits with different signs around the initial angle of the nearly closed orbits found in step 1. (3) Since we have launched trajectories in the radial direction, the accurately closed orbits must return to the spherical surface perpendicularly, which means that $p_\theta = 0$. We search these accurately closed orbits by iteration. The codes compute the relevant classical quantities in the last loop of iteration.

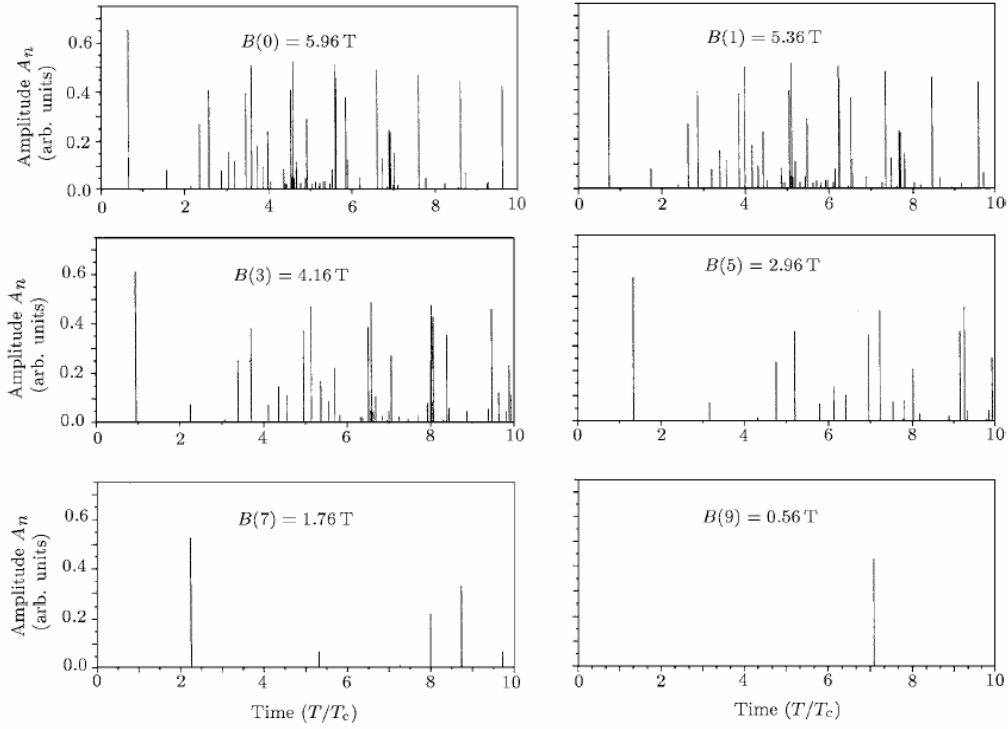

**Fig. 3.** Predicted amplitudes of oscillations in various magnetic fields. We see that some of the closed orbits disappear and those which do not disappear shift to the right and the intensities become smaller when $B(t)$ decreases. $T_c$ is the cyclotron period of the electron when $B(t) = 5.96$ T (see text).

It is well known that one of the characteristics of the system which shows classical chaos is that the evolution of the system is sensitive to its initial conditions. We show this very intuitively in Fig. 1, in which we have drawn a number of trajectories whose initial launching angles $\theta_i$ are centred around $25.70°$. We have intentionally chosen $\theta_i$ to be close to that belonging to the longest transit time of all the closed orbits found for $B(t) = 5.96$ T. We see clearly that all the classical trajectories almost overlap each other when the evolution time is small enough. However, they separate more and more as the evolution time becomes longer. Most of them do not form a closed orbits when they return to the vicinity of the nucleus, and therefore do not contribute much to the oscillation spectra.

We find all the closed orbits for various magnetic fields. We calculate the oscillator strength according to Eqs.(3)–(5). The spectra are shown in Fig. 2. We see that the structure of the spectra becomes far simpler when the magnitude of the magnetic field decreases, i.e. the frequency components become fewer. We know from the closed orbit theory that there must be fewer closed orbits in the system when the magnetic field decreases. It is easy to understand the underlying physical reason for this. As the magnetic force becomes smaller, the electron can travel much further away from the nucleus before it is turned back by the magnetic field. We keep the corresponding experimental resolution fixed during the computation. The longest transit time of the electron is chosen to be $T_{max} = 10T_c$, where $T_c$, equal to $6.0 \times 10^{-12}$ s is the cyclotron period of the electron when $B(t) = 5.96$ T.

Table 1. Comparison of classical quantities for the most stable closed orbit in different magnetic fields.

|  | $B(t)$(T) | $T_n(T_c)$ | $S_2^n$(a.u.) | $A_3^n$(a.u.) |
|---|---|---|---|---|
|  | 5.96 | 0.6648 | 207.98 | 21.25 |
|  | 5.36 | 0.7394 | 216.91 | 20.84 |
|  | 4.76 | 0.8237 | 227.28 | 20.40 |
| $\theta_i = 90°$ | 4.16 | 0.9530 | 239.56 | 19.91 |
|  | 3.56 | 1.1139 | 254.46 | 19.36 |
| $\theta_f = 90°$ | 2.96 | 1.3400 | 237.14 | 18.73 |
|  | 2.36 | 1.6810 | 297.90 | 17.99 |
|  | 1.76 | 2.2545 | 332.39 | 17.08 |
|  | 1.16 | 3.4213 | 387.91 | 15.88 |
|  | 0.56 | 7.0884 | 505.48 | 13.99 |

This can be displayed more clearly by drawing the predicted magnitudes of the spectral oscillation amplitude $A_n$ versus $T_n$, as shown in Fig. 3. We notice that some of the closed orbits disappear gradually. For those that remain, their magnitudes decrease and the periods increase as the magnetic field decreases. This can be accounted for by the decrease of the stability for the closed orbit. We can see this clearly from Table 1, in which the classical quantities of the most stable closed orbit ($\theta_i = 90°$ are listed, one of which is perpendicular to $B(t)$) in various magnetic fields. The period becomes longer and the classical three-dimensional amplitude becomes smaller as $B(t)$ decreases, and apparently $S_2^n$ increases. Generally, we find that the closed orbit with the same initial angle will be topologically identical only to different extensions in space and the relevant classical quantities change in the same manner with that of the most stable orbit ($\theta_i = 90.0°$).

In conclusion, we have calculated the quantum spectra of hydrogen atoms in various magnetic fields. We have shown schematically that the closed orbits disappear gradually with the decrease of the magnitude of the magnetic field. We think this is a good way to control the closed orbits in the corresponding classical system by changing the magnetic field, and thus changing the peaks in the Fourier transform of the quantum spectra. Our experiment group will conduct corresponding experiments in the near future using a tunable superconductor magnet.